\documentclass[10pt]{article}
\usepackage{amsmath,amssymb,graphicx}
\batchmode

\newtheorem{lem}{Lemma}
\newtheorem{thm}{Theorem}

\newtheorem{cor}{Corollary}

\newcommand{\pr}{\noindent{\bf Proof}. }
\newcommand{\re}{\noindent{\bf Remark}. }
\newcommand{\res}{\noindent{\bf Remarks}. }

\newcommand{\pa}{\partial}

\newcommand{\al}{\alpha}
\newcommand{\De}{\Delta}
\newcommand{\de}{\delta}
\newcommand{\ga}{\gamma}
\newcommand{\Ga}{\Gamma}
\newcommand{\ka}{\kappa}

\newcommand{\Om}{\Omega}

\newcommand{\ep}{\epsilon}

\newcommand{\Up}{\Upsilon}

\newcommand{\cA}{{\cal A}}

\newcommand{\cD}{{\cal D}}

\newcommand{\cH}{{\cal H}}

\newcommand{\cF}{{\cal F}}

\newcommand{\cN}{{\cal N}}

\newcommand{\cJ}{{\cal J}}

\newcommand{\bbR}{{\mathbb{R}}}
\newcommand{\bbZ}{{\mathbb{Z}}}
\newcommand{\bbC}{{\mathbb{C}}}

\begin{document}

\newpage
\title{More transition amplitudes on the Riemann sphere}
\author{ 
J. Dimock\\
Dept. of Mathematics \\
SUNY at Buffalo \\
Buffalo, NY 14260 }
\maketitle

\begin{abstract}
We  consider a  conformal field theory  for bosons on the Riemann sphere.  Correlation functions 
 are defined   as singular  limits of  functional integrals.  The main result is that
these amplitudes  define transition amplitudes,
that is   multilinear Hilbert-Schmidt   functionals  on a fixed Hilbert space.
 \end{abstract} 

\section{Introduction}

We  consider  a    bosonic   conformal field theory    
on the   Riemann sphere  $\bbC_{\infty}= \bbC  \cup \{ \infty\}$.   The   theory  is  determined by 
correlation functions which are formally   defined as follows.   Let  $\ga $  a metric  on  $\bbC_{\infty}$
of the form    $\ga =  \rho  |dz|^2$.   Let   the fields  $X(z, \bar z)$ be Gaussian random variables  
with covariance which is the kernel of      $2\pi (-\De_{\ga})^{-1}$   where      $\De_{\ga}$  is the Laplacian for this   metric.
Consider the derivatives    $\pa^m X(z)$  of order   $m=1,2,3, \dots$.   Then the correlations are 
the expectations 
  \begin{equation}  \label{first}
  <  \pa^{m_1}X (z_1)   \cdots   \pa^{m_n} X(z_n)  >
  \end{equation}
This can  be  made  precise  if   the fields   are  regarded as  distributions.      If  one avoids  
coinciding  points   the   correlation functions   are also well-defined  at  sharp points.

 Our  goal in  this paper  is to   give  a precise 
 definition of the Hilbert space structure associated  with  such correlation functions. 
 This comes in two stages
 
 In the first stage (sections  \ref{two} -  \ref{six})  we   consider fields  localized  in   the  disc   $D =  \{ z:   |z|  <1  \}$.  
We  show that   associated     correlation functions   have a positivity  property under reflections through  the circle  $|z| =1$.   The  reflection    positivity  leads to the  construction of  a  Hilbert 
 space  $\cH$ by  standard methods.    We develop the structure of this Hilbert space at some length, and in 
 particular identify it as a Fock space.

 In the second stage  (section  \ref{seven})   we consider   a family 
 of  disjoint  discs  $\{D_k \} $ in    $\bbC_{\infty}$   and    correlation    functions  
 with fields  in the discs.     We   show via parametrizations  $D \to  D_k$  that  the correlation functions    define   a multilinear  functionals  on   $\cH$, i.e.  they determine
 transition amplitudes.   In fact   we obtain the stronger result that they are Hilbert-Schmidt 
 functionals.  This is the main result.

 In an earlier paper  
 \cite{Dim07}    we   considered this problem  for  fields       $[e^{ikX(z, \bar z)}]_r$,
 a regularization of  $ e^{ikX(z, \bar z)}$.   In this case   similar  but substantially weaker results were obtained.
  
  Our results  are  likely to be useful for establishing  sewing  properties.
  For this one would want to    parametrize    some of the discs   by  $D' =   \{ z:  |z|>1 \}$
 rather than  $D$.   For   progress in this direction    see
\cite{Dim04},  \cite{Dim07},  \cite{Pic07}.

This work is  also   pointing toward establishing similar results in string  theory,   i.e 
defining  string theory scattering amplitudes   as multilinear functionals on
a full Fock space.    At the end of the paper  we comment 
on the work to be done to accomplish this goal.

\section{The basic construction}  \label{two}

\subsection{}
In this section we  define the model and verify that it has the standard features of 
a conformal field theory.    Axiomatic  treatments of conformal field theory can be 
found in   \cite{FFK89},   \cite{Gaw99}, \cite{Sch97}.

We  generally  work  in  the subset  $\bbC \subset  \bbC_{\infty}$   with standard coordinates.
These    are  denoted  $x =  (x^1,x^2)$  or     $z= x^1 + ix^2,  \bar z  = x^1-ix^2$.  
We   consider   conformal   metrics  
\begin{equation}
\ga  = \rho |dz|^2=    \rho ((dx^1)^2 +  (dx^2)^2)
\end{equation}  
For   such   metrics       
\begin{equation}
|\ga(x)|  \equiv    \det \{  \ga_{ab}(x)  \}  =  \rho(x)^2
\end{equation}
and we   have the  associated volume form or   measure 
\begin{equation}
|\ga(z)|^{1/2}\   \frac{i}{2} \  dz  \wedge  d \bar z         =     |\ga(x)|^{1/2} dx  =  \rho(x)  dx  
\end{equation} 

  Derivatives are denoted   $\pa_a  =  \pa/ \pa  x^a$   and      $\pa   =   \frac 12  ( \pa_1  -  i \pa_2) $  
 and    $ \bar    \pa    =   \frac 12  ( \pa_1  +  i \pa_2)  $.  The Laplacian is 
\begin{equation}  
\De_{\ga}  =\frac{4}{ \rho}\  \bar \pa  \pa   =   \frac{1}{\rho}  \left(  \pa_1^2  +  \pa_2^2  \right)
\end{equation}
The  Laplacian is self-adjoint  on  the  complex Hilbert space   $L^2( \bbC,   \rho \  dx   )$.
The  Laplacian  vanishes on the constants and we  consider the subspace  $\{1\}^{\perp}$
of functions  orthogonal to constants,  that is functions for which  $ \int   f(x)   \rho(x)  dx
=0$.    The Laplacian is invertible on this subspace.   The inverse can be related to the inverse for
the flat metric  and   for   $f,h \in  \{1\}^{\perp}$  one finds
\begin{equation}  \label{bingo}
2 \pi  (f,(  - \De_{\ga} )^{-1}h )  =    \int   \overline{  f(x)}  \rho(x) \log |x-x'|  \rho(x')   h(x')  dx  dx'
\end{equation}

Now  let    $X(f)$  be a   family  of Gaussian random  variables  indexed  by real  functions   $f   \in     \{1\}^{\perp}$
  with mean zero and covariance   $2\pi (  - \De_{\ga})^{-1}$.     Expectations are  denoted
by  $< \cdot >_{\ga}$   so we have   
the two point function  
\begin{equation}
<X(f)X(h)> _{\ga}  =  2 \pi(f,(  - \De_{\ga} )^{-1}h )  
\end{equation}

The correlation functions for   $X(f)$  satisfy a basic reflection positivity property.    Let   $\theta  $  be  reflection through the 
circle  $|z|=1$,  i.e.  $\theta (z)  =  \bar z^{-1}$.   Suppose that   the metric    $\ga  =\rho|dz|^2$   is chosen so   reflection is an isometry.  i.e   $\rho$ must satisfy    
   $ \rho(z)  =  |z|^{-4}\rho(\bar z^{-1})$.  Then  $\theta $ induces an anti-unitary     map $\Theta$  on the underlying   complex  $L^2$  space   such  that 
\begin{equation}  \label{ringring}
\Theta  ( X(f_1)  \dots  X(f_n) )  =   X(\theta^* f_1 )  \dots  X(\theta^*  f_n) 
\end{equation}
where  $\theta^* f  =  f \circ  \theta$.
Then   if  $F$  is any polynomial in   $X(f)$  or more generally any  $L^2$ function we have the positivity
\begin{equation}  \label{quoted}
<  \Theta (F) F  >_{\ga} \  \geq\  0
\end{equation}
This is proved in  \cite{Dim07}  by  using an approximation by massive fields and a Markov property.
An  alternative  would be to give a   basic  Fock  space  construction  of   the correlation functions in which case  the
positivity would be obvious.  (We  obtain Fock space as a derived object later on.)

\subsection{}

Now we   want   to   consider    fields   at    sharp points.      The field
$X(z,\bar z )$  is not well-defined  but the derivatives    $\pa  X(z)$   are as we now explain

A   delta  function  at  $z \in \bbC$   is the  distribution which sends the test   function  $f$  to    $<  \de_z,   f>   =  f(z)$.
We  want to approximate it by  functions on $\bbC$.  With the metric  $\ga  = \rho |dz|^2$  a   function   $h$   defines  a distribution
by    $h \mapsto   <h,f>_{\rho}$   where   
 \begin{equation}
<h,f>_{\rho}   =  \int h(x)  f(x)   \rho(x)  dx
\end{equation}
Let    $\de_{\ka}( \cdot  - z)$  be  a family   of  smooth    functions with shrinking supports   converging  to   $\delta_z$ function 
 in the flat  metric  $|dz|^2$.
Then   $  \rho^{-1}   \de_{\ka}   (\cdot -z)  $    converges to the delta  function in the metric  $ \rho |dz|^2$
for we  have  
 \begin{equation}
\lim_{\ka \to  \infty}  <  \rho^{-1}    \de_{\ka}(\cdot  -z) ,f>_{\rho}   
    =\lim_{\ka \to  \infty}  <     \de_{\ka}(\cdot  -z) ,f>_{1}              =  f(z)
\end{equation}

  The derivative of a     delta  function  at   $z \in \bbC_{\infty}$  is the map   taking  smooth   functions  $f$
 on   $\bbC_{\infty}$
 to   $ ( \pa  f ) (z)$.
 Then     $\rho^{-1}  ( - \pa )  \de_{\ka}   (\cdot -z)  $
 is an approximation   for   we  have   
   \begin{equation}
\lim_{\ka \to  \infty}  < \rho^{-1}   (-  \pa  )  \de_{\ka}(\cdot  -z) f  >_{\rho}  = 
\lim_{\ka \to  \infty}  <  \de_{\ka}(\cdot  -z) ,\pa  f  >_{1} =
\pa f(z)
\end{equation}

Thus  as  an approximation to    $\pa X(z)$   we  consider fields   $ X(\rho^{-1}(- \pa)  \de_{\ka}(\cdot -z))$. 
Note  that   the function  $\rho^{-1}(- \pa)  \de_{\ka}(\cdot -z)$  is orthogonal to constants so this is well-defined.

\begin{thm}  {  \  }    \label{looney}  For conformal  metrics   $\ga   =  \rho  |dz|^2$ 
\begin{enumerate}
\item   If  $z \neq  z'$ the limit   
\begin{equation}
C(z,z')  =   \lim_{\ka \to \infty}   
\Big<    X \Big( \rho^{-1} (- \pa)  \de_{\ka}(\cdot -z) \Big )  \   X \Big( \rho^{-1} (- \pa)   \de_{\ka}(\cdot -z') \Big )  \Big>_{\ga}
\end{equation}
exists,  is independent of  $\ga$,   and is given  by 
\begin{equation}
C(z,z')  =    - \frac{1}{2}  \    \frac{1}{(z-z')^2}
\end{equation}
\item  
For non-coinciding  points   $z_1,  \dots,  z_n$  the limit
\begin{equation} C(z_1, \dots, z_n)  =
\lim_{\ka \to \infty}   
\Big<    X\Big( \rho^{-1}  (- \pa)  \de_{\ka}(\cdot -z_1) \Big)  \cdots  X \Big(  \rho^{-1}  (- \pa)   \de_{\ka}(\cdot -z_n) \Big)  \Big>_{\ga}
\end{equation}
exists and is independent of $\ga$.   It vanishes if  $n$ is odd and if $n$ is even is given by  
 \begin{equation}  \label{pairing}
  C(z_1, \dots, z_n  )   =  \sum_{P}   \prod_{  \{\al, \beta\} \in P   }  C(z_{\al},z_{\beta})  
 \end{equation}
 where the sum is over all  pairings  $P =  \{  \{ \al_1, \beta_1 \},  \dots ,    \{ \al_{n/2}, \beta_{n/2} \}\}$  of 
 $(1, \dots,  n)$.
 \end{enumerate}
\end{thm}
\bigskip
  
 \re     The  function     $ C(z_1, \dots, z_n  ) $   gives a meaning to the formal  expression
     $ <  \pa X (z_1)   \cdots  \pa  X(z_n)  >$.  Note that  it is symmetric and analytic  away from
     coinciding points.
  \bigskip

\pr   We  have  
\begin{equation}
\pa_z   \pa_{z'}  \Big( - \log|z-z'| \Big)  =   - \frac{1}{2}  \     \frac{1}{(z-z')^2}  = C(z,z')
\end{equation}
and so   by (\ref{bingo})
  with    $d^2w  =  d (\textrm{Re }w)  \   d (\textrm{Im }w)  $\begin{equation}  \label{rrr}
\begin{split}
&\Big<    X \Big( \rho^{-1} (- \pa)  \de_{\ka}(\cdot -z) \Big )  \   X \Big( \rho^{-1} (- \pa)   \de_{\ka}(\cdot -z') \Big )  \Big>
\\
= &\int   \Big (   - \pa_{w}  \de_{\ka}(w -z)  \Big)\Big(- \log|w- w'|\Big)  \Big(   - \pa_{w'}  \de_{\ka}(w' - z')  \Big)  d^2w\  d^2w'\\
= &\int \de_{\ka}(w -z) )  C(w,  w') \de_{\ka}(w' -z')   \  d^2w\  d^2w'\\
\end{split}
\end{equation}
This   converges to  $C(z,z')$.   For the second part the regularized expression is a standard Gaussian 
integral.   It   is   evaluated as   (\ref{pairing})  with  $C(z,z')$  replaced by   
   (\ref{rrr}).  The  result  follows by taking the limit.
\bigskip

The   correlation  functions  are so far only defined in the coordinate patch  $\bbC$  and we  add some remarks on 
including the point at  infinity.    For this     we  go to the other coordinate patch  on  $\bbC_{\infty} -\{ 0  \}$  with coordinates   $\zeta  =  1/z$.    The metric is now  $\hat  \ga  = \hat  \rho  | d \zeta|^2$  where
\begin{equation}
\hat   \rho  (\zeta)   =   \rho(1/\zeta)   \frac{dz}{d \zeta}  \frac{d  \bar  z}{d   \bar\zeta}
=    \rho(1/\zeta  )  |\zeta|^{-4}
\end{equation}
With   $\hat  f(\zeta)   =   f (1/\zeta)$   the expression  (\ref{bingo})  becomes in new  coordinates
\begin{equation}  \label{bingo2}
2 \pi  (f,(  - \De_{\ga} )^{-1}h )  =    \int   \overline{  \hat f(\zeta)} \hat   \rho(\zeta )  \left(-
 \log \left|\frac{1}{\zeta}-  \frac{1}{\zeta'}   \right| \right) \hat   \rho(\zeta')  \hat h(\zeta')  d^2\zeta  d^2\zeta'
\end{equation}
Now  let   $\hat f   =   \hat  \rho ^{-1}  (-\pa)  \delta  (  \cdot - \zeta)$    and  
 $\hat h  =   \hat  \rho ^{-1}  (-\pa)  \delta  (  \cdot - \zeta')$ 
and  take  the limit   $\kappa  \to  \infty$   to  get  the two point function in these coordinates.  
We find  for  $\zeta, \zeta' \in  \bbC$
\begin{equation}
\hat  C(\zeta,  \zeta')   =  - \frac{1}{2}  \frac{1}{(\zeta-\zeta')^2}\end{equation}
Note that  in each variable   the correlation  functions transform  as one-forms for  we have  
 on  $\bbC -  \{ 0\}$  
\begin{equation}  \label{chicken}
\hat C(\zeta,\zeta')    =  - \frac{1}{2}  \frac{1}{(\zeta-\zeta')^2} =
  C(\frac{1}{\zeta},  \frac{1}{\zeta'} )  \frac{ dz}{ d \zeta} \frac{ dz'}{ d \zeta'} 
\end{equation}
Similarly  if   only  one point is expressed in new coordinates  we find   
\begin{equation} 
\tilde C(z ,\zeta'   )  =         \frac{1}{2}   \frac{1}{(1-  z\zeta')^2}  = C(z,   \frac{1}{ \zeta'} )  \frac{ dz'}{ d \zeta'} 
\end{equation}

Returning to standard coordinates
let us also note      the   transformation properties     under  general    Mobius transformations
\begin{equation} 
w =  \al(z)  =   \frac{az+b}{cz+d}
\end{equation}
and   radial  reflections.

\begin{lem}    {  \  }
\item 
For   a  Mobius transformation   $w_i  =  \al  (z_i)$
 \begin{equation}
   C( z_1, \cdots, z_n)= C( w_1, \cdots, w_n)  \prod_i  \frac{ \pa  w}{\pa z} (z_i) 
  \end{equation}
 \item  Under   radial reflections        $\theta z  =  1/ \bar z$
 \begin{equation}  \label{stun}
 \overline{   C( z_1, \cdots, z_n)}=
     C\left(\frac{1}{\bar z_1}, \cdots, \frac{1}{\bar z_n}\right)  \prod_i  \frac{-1}{\bar z_i^2}
 \end{equation}      
\end{lem}
\bigskip

\pr     It  suffices to show these  results for the two point function.   In the first case   for  $w=\al(z),  w'= \al(z')$
 \begin{equation}
   C( z,z')=  \frac{ \pa  w}{\pa z}  \frac{ \pa  w'}{\pa z'}  C(w,w') 
  \end{equation}
If   it holds   for   Mobius  transformations   $\al, \beta$  then it holds  for the composition  $\al \circ  \beta$.
But any  Mobius is a composition of translations,  scalings,  and inversions.    Thus it suffices to 
check for each of these separately.   This is easy  for   $w=z+b$ and  $w = q z$  and for     $w=1/z$
we have have  already checked it.    

For reflections we have    
\begin{equation}    \label{whale}
 C\left(\frac{1}{\bar z}, \frac{1}{\bar z'}\right) \frac{-1}{\bar z^2  }   \frac{-1}{\bar z'^2}    = - \frac12\ \frac{1}{(\bar  z-  \bar z')^2}  = \overline{  C(z,z')  }
\end{equation}
This completes the proof.

\subsection{}
We also need  a reflection positivity property.  To formulate it   and for  future  purposes  we 
reformulate our results in a more algebraic  language  as  in   \cite{Dim07}.

 Let  $\Up$  be the free algebra generated by  the sequences  $Z=   [z_1, \dots, z_n]$  with   $z_i  \in\bbC$. 
 We  include the empty sequence.
The general  element  
is a finite sum     
\begin{equation}
F  = \sum_Z  F(Z)  Z  \hspace{1cm}   F(Z) \in \bbC
\end{equation}  
 and   multiplication  $ZZ'$   is   juxtaposition.  Let  $\Up_0$  be the subspace   
 spanned by  elements $Z$  with non-coinciding points.   
 We  define   a  linear function  on  $F  \in  \Up_0$ by     
 \begin{equation}
<F>  =   \sum_Z  F(Z)  <Z>
\end{equation}
where   
 \begin{equation} 
 <Z>   =  <  [z_1, \dots,  z_n]>=   C(z_1, \dots,  z_n)
 \end{equation}
 Thus   $ [z_1, \dots,  z_n]$  is a symbol  for  the  formal random variable      $   \pa X (z_1)   \cdots  \pa  X(z_n) $. 
  In table  1
 we   list   the  conventions of this type that we employ.

\begin{table}   
\hspace{3cm}\begin{tabular}{||l r | l ||}  \hline
formal random variable  &    & symbol   \\  \hline  \hline
& &  \\
$\pa X(z_1) \dots  \pa X(z_n)$  &   &    $[z_1, \dots, z_n]  \in  \Up$   \\  
& &  \\
\hline
& &  \\
$\pa ^{m_1}X(z_1) \dots  \pa^{m_n} X(z_n)$  &   &    $[m_1, z_1, \dots,m_n,  z_n]  \in   \Up'$   \\ 
& &  \\ 
\hline
& &  \\
$:\pa ^{m_1}X(z_1) \dots  \pa^{m_n} X(z_n):$  &   &    $:[m_1, z_1, \dots ,m_n,  z_n]:_0 \  \in  \Up'$  \\ 
& &  \\ 
\hline
& &  \\
$:\pa ^{m_1}X(z_1) \dots  \pa^{m_n} X(z_n):$  &   &    $:[m_1, z_1, \dots,m_n,  z_n]:  \ \in   \Up^w $ \\ 
& &  \\ 
\hline
\end{tabular}
\caption{  Formal random variables are represented by symbols which are 
well-defined elements of the   various algebras   $ \Up,  \Up', \Up^w$  }
\end{table}

Now   define radial reflection on symbols    
  $Z = [z_1, \dots,  z_n]$  with  $z_i \neq  0$  by   
  \begin{equation}
\Theta  Z  =  \Theta   [  z_1,  \dots,   z_n]
=    \prod_{i=1}^n   \frac{-1}{\bar z_i^2} \left  [ \frac{1}{\bar z_1}, \cdots, \frac{1}{\bar z_n}\right ]
 \end{equation} 
 Then   for non-coinciding  points we have by  (\ref{stun})  that     $\overline { <Z> } =  <  \Theta  Z>$.  
 We   extend $\Theta$   to the subspace  of  $\Up$   spanned by symbols with  $z_i \neq  0$  by  defining for  $F  = \sum_Z  F(Z)  Z $
\begin{equation}
\Theta  F   =  \sum_Z   \overline  {F(Z)} \Theta Z
\end{equation}
Then    for non-coinciding points  $\overline { <F> } =  <  \Theta  F>$.    The operator    
  $\Theta $  is   an anti-linear automorphism satisfying  $\Theta^2 =1$.

Next  let   $D, D'$   be  half-spheres    (discs)
\begin{equation}
D   =  \{  z \in \bbC :    |z|  <1  \}  \hspace{1 cm}   D'   =  \{  z \in \bbC:    |z|  >1  \}  \hspace{1 cm}   
\end{equation}
We  are particularly interested in  symbols  $Z=   [z_1, \dots, z_n]$  with  non-coinciding points
and    $z_i \in D-\{0\}  $.   These  span   a subspace of   $\Up_0$  denoted
  $\Up_{0,D-\{0\}}$.   Similarly   $\Up_{0,D'}$  has   non-coinciding points  in   $D'$. 
  Note that if    $F, G  \in \Up_{0,D-\{0\}}$  then  $\Theta F  \in   \Up_{0,D'}$  and  
$(\Theta  F) G   \in  \Up_{0}$.

\begin{lem}    \label{reflection}  For  $F,G  \in   \Up_{0,D-\{ 0\}}$
\begin{equation}
   <  (\Theta F) G> 
=   \overline {<  (\Theta G) F>}
\end{equation}
and  we have the  reflection positivity
  \begin{equation}     
<  (\Theta  F)  F >  \ \geq   0   
\end{equation}
\end{lem}
\bigskip

\pr    The  first identity    follows from  $\overline{<  F> }  =  < \Theta  F> $   and    $\Theta^2 =1$  and 
  the fact that $\Theta$ is an automorphism.

For the positivity we have  to  show that   
 \begin{equation}
 \sum_{Z,W}   \overline{ F(Z)  } F(W)    < ( \Theta Z) W >\   \geq  0
  \end{equation}
Choosing a    reflection invariant metric   $\ga = \rho|dz|^2$ we have   for    $Z = [z_1, \dots,  z_n]$ and
$W= [w_1, \dots,  w_m]$
\begin{equation}
\begin{split}
 < ( \Theta Z) W >   =& \prod_i   \frac{-1}{\bar z_i^2}    <  [ \bar  z_1^{-1}, \dots,  \bar   z_n^{-1},w_1,  \dots,  w_m] >\\
=&   \lim_{\ka \to \infty}  \left<  \prod_{i=1}^n  X  \Big(  \frac{-1}{\bar z_i^2}  \rho^{-1} ( -  \pa )\de_{\ka}( \cdot -  \bar  z_i^{-1}) 
\Big)\  \prod_{j=1}^m  
 X\Big(   \rho^{-1} ( -  \pa )\de_{\ka}( \cdot - w_j) \Big)\right>_{\ga}\\
\end{split}
 \end{equation}  
 The  reason this works is that  
 $  <\rho^{-1} ( -  \pa )\de_{\ka}( \cdot -  \bar  z^{-1}),f>_{\rho} $   converges to     $    \pa f ( \bar z^{-1})$. 
  But  we  can replace  $\rho^{-1}( -  \pa )\de_{\ka}( \cdot -  \bar  z^{-1})$  with another approximating
  sequence  and  we  take 
    $- \bar z^2  \theta^* (\rho^{-1}(-  \bar  \pa)  \delta_{\ka}  ( \cdot  -  z  ) )$.
      This works since  $\theta$ is an isometry and so  
    \begin{equation}
\begin{split}
  < \theta^* (\rho^{-1}(-  \bar  \pa)  \delta_{\ka}  ( \cdot  - z  ),f >_{\rho}  
=  &    <\rho^{-1}(-  \bar  \pa)  \delta_{\ka}  ( \cdot  -  z  ), \theta^*  f >_{\rho}  \\
= &     \int     \de_{\ka}( w - z)  \pa_{\bar w}   f(  \bar w^{-1}  )dz \\
=&     \int    \de_{\ka}( w-z)  ( -\bar w^{-2} )   \pa f(  \bar w^{-1}  ) ) dz \\
\end{split}
\end{equation}
which   converges to    $-  \bar z^{-2}     \pa f(  \bar z^{-1}  ) $. 
Thus we have   
  \begin{equation}
\begin{split}
 \Big< ( \Theta Z) W  \Big >   
=&   \lim_{\ka \to \infty}   \left<  \prod_{i=1}^n  X  \Big(  \theta^* (\rho^{-1}(-  \bar  \pa)  \delta_{\ka}  ( \cdot  -  z_i  ) )
\Big)\  \prod_{j=1}^m  
 X\Big(   \rho^{-1} ( -  \pa )\de_{\ka}( \cdot - w_j) \Big)\right >_{\ga}\\
 =&   \lim_{\ka \to \infty}  \left < \Theta  \left( \prod_{i=1}^n  X  \Big(  (\rho^{-1}(-    \pa)  \delta_{\ka}  ( \cdot  -  z_i  ) )
\Big)  \right)    \  \prod_{j=1}^m  
 X\Big(   \rho^{-1} ( -  \pa )\de_{\ka}( \cdot - w_j) \Big)\right >_{\ga}\\
\end{split}
 \end{equation}  
where    now  $\Theta$  is the operator on random variables defined in (\ref{ringring}).    The positivity 
for   $\ka  < \infty$ 
now follows by   the positivity result for random variables  quoted in  (\ref{quoted}),  and hence it also holds in the limit.
 \bigskip

\subsection{}

The reflection positivity   gives   a Hilbert space structure  as  follows.  
Let  $\cN$ be the null space  all  $F$ 
in    $\Up_{0,D-\{0\}}$     such that       $< (\Theta F) F>   =0$.    By theorem  \ref{reflection} 
  $< (\Theta F) G>   $
determines an inner  product on    $\Up_{0,D-\{0\}}/  \cN$.     If   $\nu_0$   is     the mapping from   $\Up_{0,D-\{0\}}$  to its equivalence class in    $\Up_{0,D-\{0\}}/  \cN$   then    $(\nu_0(F),  \nu_0(G))   =  < (\Theta F) G>  $.      The completion in the norm  $\|\nu_0(F)\|  =   < (\Theta F) F>^{1/2}    $   is    then   a Hilbert  space
\begin{equation}
\cH    =  \overline{   \Up_{0,D-\{0\}}/ \cN  }
\end{equation}  Thus we have

\begin{thm}  {  \  }  \label{recon} There exists a complex   Hilbert  space  $\cH$ 
 and a linear  mapping  $\nu_0:  \Up_{0,D-\{0\}}   \to  \cH$
with dense range   such that    for all  $F,G  \in    \Up_{0,D-\{0\}} $
\begin{equation}
( \nu_0(F),   \nu_0(G)   )  =  <   ( \Theta  F)G  >
\end{equation}
\end{thm}
\bigskip  

\begin{cor}   \label{snug} 
  The  function   $\nu_0([z_1, \dots,  z_n])$    is   strongly analytic   $z_i  \in  D-\{ 0\}$    away from coinciding  points.
  It is symmetric under permutations of the $z_i$.
\end{cor}

\pr   The  expression 
\begin{equation}
(  \nu_0([w_1, \dots,  w_r]),    \nu_0([z_1, \dots,  z_n]))
=    \prod_i   \frac{-1}{\bar w_i^2}    <  [ \bar  w_1^{-1}, \dots,  \bar   w_r^{-1},z_1,  \dots,  z_n] >
\end{equation}
is  analytic  in   $z_i  \in  D-\{ 0\}$  away from coinciding  points since the $\bar  w_j^{-1}$ are in  $D'$
and the correlation functions have this property.  Hence   $(  \nu_0(F),    \nu_0([z_1, \dots,  z_n]))$ is 
analytic  for  all    $F \in    \Up_{0,D-\{0\}}$.   The  $ \nu_0(F)$  are a dense   set  so  $  \nu_0([z_1, \dots,  z_n])$  is
weakly analytic  and hence strongly  analytic.  The symmetry follows similarly.
\bigskip

\re   One can show  that     For   $z  \in   D-\{0\}$  there exists  an operator     $\pa  \hat  X(z)$
defined   on   the span of    $\nu_0([z_1, \dots,  z_n])$    with   $|z_i|  < |z|$ such that   
\begin{equation}  
\pa  \hat  X(z)\nu_0 ([z_1, \dots, z_n])   =   \nu_0 ([z, z_1, \dots, z_n])
\end{equation}
 If  we let   $\Om  =  \nu_0(\emptyset)$   then  we have    
 for   $0< |z_n| <  \cdots  <  |z_1| < 1$ 
\begin{equation}
 \nu_0 ( [ z_1, \dots,  z_n] )  =    \pa \hat X(z_1)  \cdots  \pa  \hat  X(z_n)  \Om
 \end{equation}
We   will   not   need  these  field operators  directly,   see  Gawedski  \cite{Gaw99}  for details.  
We  will  however   treat the closely related creation and annihilation operators.

\subsection{}   \label{serious}
We   can make a similar construction based  on  any  disc
in $\bbC$,     see also  \cite{FFK89}.  
Consider  a disc of the form 
\begin{equation}
\tilde D   =   \{z \in \bbC:   |z-a|< R  \} 
\end{equation}
Radial  reflection  is now defined by  
\begin{equation} 
\tilde  \theta   (a +z)   = a + R^2/  \bar z
\end{equation}
An  anti-linear  automorphism      $\tilde  \Theta$   on    $ \Up_{0, \bbC-  \{ a\}}$     is defined by 
\begin{equation}  
\tilde  \Theta   [a+z_1, \dots,  a + z_n]   
=   \prod_i  (  -  R^2/\bar z_i^2 ) \     [a+R^2/\bar  z_1, \dots,  a + R^2/\bar  z_n]   
\end{equation}
This  satisfies  $\tilde \Theta^2 = I$   and    and for   $F,G  \in  \Up_{0,\tilde  D-\{a\}}$
\begin{equation}
  <  (\Theta F) G> 
=   \overline {<  (\Theta G) F>}
     \hspace{2cm}     < ( \Theta  F)   F> \     \geq   0  
\end{equation}
We  then create a Hilbert  space   
\begin{equation}  \label{fifty}
\cH_{\tilde D}      =\overline{  \Up_{0,\tilde D-\{a\}} /  \cN}
\end{equation}
as before.  If  $\tilde \nu $  maps to equivalence  classes then
  $(\tilde  \nu(F), \tilde   \nu(G)   )  =  <   ( \Theta  F)G  >$.

We   identify the spaces    $\cH$, $\cH_{\tilde D} $ as follows.     For  any  $a  \in \bbC,  q  \in \bbC-\{0\}$ 
   define  an isomorphism $J$   on 
$\Up$ by 
\begin{equation}
J[z_1, \dots, z_n]   =    q^n      [a +qz_1, \dots,a+  qz_n] 
\end{equation}
This is normalized so that   $<JZ>=<Z>$   and hence   $<JF>  =   <F>$ for   $F  \in  \Up_0$. 
If   $|q| =R$  then   $J$ maps  $\Up_{0,D-\{0\}}$
 onto $\Up_{0,\tilde D-\{a\}}$
and satisfies  $\tilde \Theta  J   =  J \Theta$.   Furthermore  $J$   preserves inner 
products  since   for   $F,G  \in  \Up_{0, D-\{0\}}$ 
\begin{equation}
    < (\tilde   \Theta J F) J  G>
        =    < (J  \Theta  F) J  G>
  =        < (   \Theta  F)   G> 
 \end{equation}
Hence   $J$   determines a unitary operator  $\cJ:   \cH  \mapsto  \cH_{\tilde D}$   such that  
    \begin{equation}
    \cJ   \nu_0   ( F)   =  \tilde   \nu  (J  F)
 \end{equation}        
        
There is still   some arbitrariness  associated with the choice of  $q$  with $|q| =R$.  Making
a choice  of  $q$   is choosing a parametrization of  $\tilde D$.

\section{More derivatives}

\subsection{}
Now    we    add  more  derivatives.   Since   $C(z_1, \dots, z_n)$  is analytic  away from 
coinciding points we can define symmetric analytic functions  by  
\begin{equation}
C(m_1,z_1, \dots,m_n, z_n)  = \pa_{z_1}^{m_1-1}  \cdots  \pa_{z_n}^{m_n-1}C(z_1, \dots, z_n)
\end{equation}
This gives a meaning to the formal expression   $<\pa ^{m_1}X(z_1) \dots  \pa^{m_n} X(z_n) >$.
By   (\ref{pairing})  this is evaluated as 
 \begin{equation}
 C(m_1,z_1, \dots,m_n, z_n)   =     \sum_{P}   \prod_{  \{\al, \beta\} \in P   } 
  C(m_{\al},z_{\al},m_{\beta},z_{\beta}) 
   \end{equation}
where
\begin{equation}  \label{lady}
C(m_1,z_1,m_2,z_2)   =   - \frac{1}{2}\ \pa^{m_1-1}_{z_1} \pa^{m_2-1}_{z_2}  \frac   {1} {  (z_1 -z_2)^2    } 
=     \frac{1}{2}\  \frac{     ( m_1+m_2 -1)!  (-1)^{m_1}    }{     (z_1 -z_2)^{m_1 + m_2} }        
\end{equation}
These correlation functions  can also be obtained as limits of more regular objects   as in theorem  \ref{looney}.

Mobius transformation have a complicated effect  on these correlations.  (The $\pa^mX(z)$ are not primary   fields).
Still there is  a simple scaling property.  For  $q, a  \in  \bbC$ 
\begin{equation}  \label{scaling}
 C ( m_1,z_1, \dots, m_n,z_n)
=    C(m_1,qz_1+a, \dots, m_n,qz_n+  a)
\prod_i q^{ m_i}
  \end{equation}

Radial  reflections   also have a complicated effect,  but in this case  we need an explicit 
expression.    We   still expect the effect on correlation functions to be complex conjugation  so 
to  find a  candidate we  compute
 \begin{equation}  \label{grape}
\begin{split}                                                                        
 \overline { C(m_1,z_1, \dots,m_n, z_n)  }
   =   &      \pa^{m_1-1}_{\bar  z_1}  \cdots   \pa^{m_n-1} _{\bar  z_n}  \overline{ C(z_1, \dots, z_n)  } \\
=& \pa^{m_1-1}_{\bar  z_1}  \cdots   \pa^{m_n-1} _{\bar  z_n} 
 \left(     \prod_j -\bar z_j^{-2}\  
 C( \bar z_1^{-1},  \dots,   \bar    z_n^{-1}) \right)  \\
=&     \sum_{a_1, \dots,  a_n} 
\Big( \prod_j  d_{m_j,a_j}      \bar z_j^{- 2 -m_j-a_j}  \Big)
 C(a_1, \bar  z_1^{-1}, \dots, a_n,  \bar z_n^{-1}) \\
\end{split}
\end{equation}
Here  in the last  line  we have  
defined       
 integers     $d_{m,a}$  by  computing for any analytic function $f$
 \begin{equation}
\bar  \pa^{m-1}  \Big(-  \bar z^{-2}\  \pa f(  \bar z^{-1}  )\Big)
=   \sum_{a=0}^{m} d_{m,a} \   \bar z^{- 2 -m-a}\    \pa^a  f  (\bar z^{-1})
\end{equation}
or equivalently 
\begin{equation}
  \bar  \pa^{m}  \Big( f  (  \bar z^{-1}  )   \big)
=   \sum_{a=0}^{m} d_{m,a} \   \bar z^{- 2 -m-a}\    \pa^a  f  (\bar z^{-1})
\end{equation}

\subsection{}

We   want to  work the derivatives       into our basic Hilbert space
structure. 
Let       $\Up'$   be the free algebra generated by   the sequences   $Z= [m_1,z_1, \dots, m_n,z_n]$
with  $m_i \geq  1$  and  $ z_i  \in \bbC$.
We identify   $\Up$  as the subalgebra  generated by  $ [1,z_1, \dots, 1,z_n]$.
 Let  
$\Up'_{0}$  elements of  $\Up'$   with non-coinciding points.  We  extend our expectation   on $\Up_0$
to  an  expectation on   $\Up'_0$   given  by    
\begin{equation}
<Z>   = <[m_1,z_1, \dots, m_n,z_n]>  =     C(m_1,z_1, \dots,m_n, z_n)  
\end{equation}
Thus  $ [m_1,z_1, \dots, m_n,z_n]$   stands for   the formal  $ \pa ^{m_1}X(z_1) \dots  \pa^{m_n} X(z_n) $.

For each  $a \in \bbC,   q  \in \bbC-\{0\}$   we  define an  define   an extended automorphism   $J$  on 
$\Up'$   by   
\begin{equation}  \label{sss}
J   [m_1,z_1, \dots, m_n,z_n]  =    [m_1,a + qz_1, \dots, m_n,a+qz_n]  \prod_i q^{m_i}
\end{equation}
 Then  by  (\ref{scaling}) we have  $<JZ>  =  <Z>$.

Taking   a  cue   from  (\ref{grape})  we define an  extended reflection  as     an  anti-linear    automorphism   on  $\Up'_{\bbC-\{0\}}$
by  
\begin{equation}    \label{all}
\Theta    [m_1,z_1, \dots, m_n, z_n]
=    \sum_{a_1, \dots,  a_n} 
\Big( \prod_j  d_{m_j,a_j}      \bar z^{- 2 -m_j-a_j}  \Big)
 [ a_1, \bar  z_1^{-1}, \dots, a_n,  \bar z_n^{-1}]
 \end{equation}
Then    $\overline{<  Z> }  =  < \Theta  Z> $  .  

\newpage

\begin{lem}  {  \  }  \label{sweden}
  $\Theta^2  =  I   $ 
\end{lem}
\bigskip

\pr   This follows from the identity 
\begin{equation}
\sum_{0  \leq  b  \leq   a  \leq  m}    d_{m,a}   d_{a,b}   =  \de_{m,b}
\end{equation}
To see this   let      $g( z)  =  f(1/z)$    and then by definition  
 \begin{equation}    \label{fir}
  (  \pa^a  g )( z)   = \sum_{0\leq  b \leq  a}     d_{a,b} \   z^{-2-a-b}  \  (\pa^b f) (1/  z  ) 
  \end{equation}
On the other hand since      $f(z)  =  g(1/  z)$ 
 \begin{equation}     \label{sec}
  ( \pa^m  f )( z)   = \sum_{0\leq  a \leq  m}     d_{m,a} \    z^{-2-m-a}  \  (  \pa^a g) (1/  z  ) 
  \end{equation}
 Substitute    (\ref{fir})  at   $1/z$    into  (\ref{sec})   and obtain 
  \begin{equation}    
  ( \pa^m  f )( z)   = \sum_{0\leq b \leq    a \leq  m}     d_{m,a} d_{a,b}   z^{b-m}  \  (   \pa^b f) (z  ) 
  \end{equation}
Now  take   $f(z)  = e^z$   and get   
  \begin{equation}
  z^m   =    \sum_{0\leq b  \leq  m} \left( \sum_{ b \leq    a \leq  m}       d_{m,a} d_{a,b}  \right) z^b  \  
  \end{equation}
  Matching coefficients we have the desired result. 
\bigskip

Now  let   $\Up'_{0, D- \{0\}}$   be  elements with non-coinciding points  in  $D-\{0\}$
and  extend the definition of  $\nu_0:  \Up_{0,D-\{ 0\}}  \mapsto  \cH$   to   a  linear  map    
 $\nu_0:  \Up'_{0,D-\{ 0\}}  \mapsto  \cH$  by  
\begin{equation}
\nu_0(Z)  \equiv   \nu_0( [m_1,z_1, \dots  m_n, z_n])   =     \pa^{m_1-1}_{z_1}  \cdots   \pa^{m_n-1} _{z_n}  \nu_0 ([z_1, \dots, z_n])
  \end{equation}

\begin{lem}  \label{away}
For   $F,G   \in   \Up'_{0,D-\{ 0\}} $
\begin{equation}
( \nu_0(F),   \nu_0(G)   )  =  <   ( \Theta  F)G  >
\end{equation}
\end{lem}
\bigskip

\pr   
For  $Z=    [m_1,z_1, \dots  m_n, z_n]$   and   $W= [\ell_1, w_1, \dots,\ell_r, w_r]$
 \begin{equation}   \label{poopoo}
 \begin{split} 
  ( \nu_0(Z), \nu_0(W)) )
 = & \pa^{m_1-1}_{\bar  z_1}  \cdots   \pa^{m_n-1} _{\bar  z_n} 
 \pa^{\ell_1-1}_{  w_1}  \cdots   \pa^{\ell_r-1} _{  w_r} \Big( 
 \nu_0  (  [ z_1, \dots, z_n]  ),\nu_0  (  [ w_1, \dots, w_r] )\Big)\\
=& \pa^{m_1-1}_{\bar  z_1}  \cdots   \pa^{m_n-1} _{\bar  z_n} 
 \pa^{\ell_1-1}_{  w_1}  \cdots   \pa^{\ell_r-1} _{  w_r}   \prod_j (-  \bar   z_j ^{-2})
 < [\bar  z_1^{-1}, \dots,\bar z_n^{-1}, w_1, \dots, w_r]>    \\
 =&      \sum_{a_1, \cdots  a_r}  \Big(
 \prod_j d_{m_j,a_j}      \bar z_j^{- 2 -m_j-a_j}  \Big)
   <  [ a_1, \bar  z_1^{-1}, \dots, a_n,  \bar z_n^{-1},\ell_1, w_1, \dots,\ell_r, w_r]> \\
   =&   <( \Theta   Z) W>  \\
 \end{split}
\end{equation}
The general result follows.

\section{Wick products}

\subsection{}
Now   we can define   Wick   products  as  elements of  $\Up$ by 
\begin{equation}
:Z:_0 =  : [z_1, \dots, z_n]:_0  
  =    \sum_Q   \prod_{  \{\al, \beta\} \in Q   } (-   C(z_{\al},z_{\beta})  )     \prod_{\ga \notin Q}  [z_{\ga}] 
\end{equation}
where  $Q$  is a collection of pairs   from
 $(1, \dots ,n)$ and we must restrict to non-coinciding points.     More generally  in  $\Up'$
\begin{equation}  \label{shut}
:Z:_0 =  : [m_1,z_1, \dots, m_n,z_n]:_0    
  =    \sum_Q   \prod_{  \{\al, \beta\} \in Q   }   ( -  C(m_\al,z_{\al}, m_{\beta}, z_{\beta}) )    \prod_{\ga \notin Q}  [m_{\ga},z_{\ga}] 
\end{equation}
where   $C$ is defined in  (\ref{lady}).   If there is a single point then    $:[m,z]:_0  =  [m,z]$.   For  future reference we note   that  
 Wick products   satisfy the identity   
\begin{equation}  \label{ugh}
\begin{split}
&[m,z]    : [m_1,z_1, \dots,m_n, z_n]:_0 \\
=  &  : [m,z,  m_1,z_1, \dots,m_n, z_n]: _0
+   \sum_{j=1}^n   C(m,z,m_j,z_j)
 : [m_1,z_1, \dots, \widehat{m_j,z_j},  \dots,  m_n, z_n]:_0 \\
\end{split}
\end{equation}
where the hat means omit  $m_j,z_j$.

 Now  consider   several Wick products
 \begin{equation}
: Z^i:_0   =  :  [ m^i_1, z^i_1, \dots ,  m^i_{n_i} z^i_{n_i}]:_0 
  \end{equation}
    Products   $:Z^1:_0 \cdots  :Z^r:_0$  are   elements   of     $\Up'$  and  if  no points  coincide  
 we  can consider the expectation   $ < :Z^1:_0 \cdots   :Z^r:_0>$.  
     As is well known Wick  products  suppress pairings within  the dots.   Hence  
\begin{equation}  \label{gentlemen}
\begin{split}
& < :Z^1:_0 \cdots   :Z^r:_0:> 
=  \sum_{P'}   \prod_{ \{(i,\al),(j, \beta) \} \in  P'}   C(m^i_{\al},z^i_{\al},m^j_{\beta},z^j_{\beta}   ) \\
\end{split}
 \end{equation}
 where the sum is over pairings   $P'   =  \{  \{(i,\al),(j, \beta)\}  \}$  with  $i\neq j$.  
 This can be thought of a sum over graphs on  $r$ vertices  with  $n_i$ legs at the $i^{th}$ vertex.
 The  correlation function is invariant under permutation of the $Z$'s  and under permutations within
the $Z$'s.

We   collect  some properties  of  Wick products.

\begin{lem}   {  \  }  \begin{enumerate}
\item   The automorphism  $J$    defined  by   (\ref{sss}) satisfies
\begin{equation}    \label{ooo}
J  : [m_1,z_1, \dots, m_n,z_n]  :_0  =   : [m_1,a + qz_1, \dots, m_n,a+qz_n]:_0  \prod_i  q^{m_i}
\end{equation}
\item   The anti-linear  automorphism defined  by   (\ref{all})      satisfies  
\begin{equation}  \label{ppp}
\Theta   : [m_1,z_1, \dots, m_n, z_n]:_0
=    \sum_{a_1, \dots,  a_n} 
\Big( \prod_j  d_{m_j,a_j}      \bar z^{- 2 -m_j-a_j}  \Big)
 :[ a_1, \bar  z_1^{-1}, \dots, a_n,  \bar z_n^{-1}]:_0
 \end{equation}
\item    $ \nu_0 (:[z_1, \dots, z_n]:_0)$   is  analytic for  non-coinciding points    $z_i   \in   D-\{ 0\}  $
and  
\begin{equation}     \label{Wick}
  \pa^{m_1-1}_{z_1}  \cdots   \pa^{m_n-1} _{z_n}  \nu_0 (:[z_1, \dots, z_n]:_0)
=   \nu_0 (:[m_1,z_1, \dots  m_n, z_n]:_0) \\
\end{equation}    
\end{enumerate}
\end{lem}
\bigskip

\pr   Each  of these  follows from the corresponding formula without the Wick ordering.
We  check   the second in detail.   By  definition 
\begin{equation}
\Theta   : [z_1,m_1, \dots, z_n,m_n]: _0   
  =    \sum_Q   \prod_{  \{\al, \beta\} \in Q   }   ( - \overline{ C(m_\al,z_{\al}, m_{\beta}, z_{\beta})} )  \Theta
  \Big(  \prod_{\ga \notin Q}  [m_{\ga},z_{\ga}] \Big)
\end{equation}
However 
\begin{equation}
\Theta   \Big(  \prod_{\ga \notin Q}  [m_{\ga},z_{\ga}] \Big)
=    \sum_{ \{a_{\ga}\}  } 
\Big( \prod_{ \ga \notin Q}    d_{m_{\ga},a_{\ga}}      \bar z^{- 2 -m_{\ga}-a_{\ga}}  \Big)
 [ a_{\ga},  \bar z_{\ga}^{-1}]
 \end{equation}
and    by  (\ref{whale})
\begin{equation}
\begin{split}
  \overline{ C(m_\al,z_{\al}, m_{\beta}, z_{\beta})} 
  = &  \pa_{\bar z_{\al}}^{m_{\al}-1}     \pa_{\bar z_{\beta}}^{m_{\beta}-1}   
\overline {  C(z_{\al},  z_{\beta}  ) }\\
 = &  \pa_{\bar z_{\al}}^{m_{\al}-1}     \pa_{\bar z_{\beta}}^{m_{\beta}-1}   
(-\bar  z_{\al}^{-2})(  -      \bar  z_{\beta}^{-2} )    C( \bar  z_{\al}^{-1},  \bar  z_{\beta}^{-1}  )\\
=   &   \sum_{ a_{\al}, a_{\beta}  } 
   d_{m_{\al},a_{\al}}    d_{m_{\beta},a_{\beta}}      \bar z^{- 2 -m_{\al}-a_{\al}}  
  \bar z^{- 2 -m_{\beta}-a_{\beta}}   C(a_\al,\bar z_{\al}^{-1}, a_{\beta},\bar   z_{\beta}^{-1})
\end{split}
\end{equation}
Thus  
\begin{equation}
\begin{split} 
&\Theta   : [m_1,z_1, \dots,m_n, z_n]:    \\
=&   \sum_{a_1, \dots , a_n}   \Big( \prod_{j=1}^n  d_{m_j,a_j}      \bar z^{- 2 -m_j-a_j}  \Big)
  \Big  [   \sum_Q   \prod_{  \{\al, \beta\} \in Q   }(-   C(a_\al,\bar z_{\al}^{-1}, a_{\beta},\bar   z_{\beta}^{-1}))
   \prod_{\ga \notin Q}  [a_{\ga},\bar  z_{\ga}^{-1}] \Big  ]
\\
 \end{split}
 \end{equation}
The bracketed expression is identified as  $ : [a_1, \bar   z_1^{-1}, \dots,a_n,  \bar   z_n^{-1}]: $  to complete the proof.

\subsection{} 
Next      want to  extend   our  results to include   coinciding  points within the 
Wick products.      The  correlation functions   are
defined for  such objects, but they are not in our algebra.   So we first enlarge the algebra.

Let   $\Up^w $   be   the free algebra  generated  by (non-empty)  symbols   
\begin{equation}
:Z:  =   :[m_1,z_1, \dots m_n,z_n]:   
\end{equation}  
with   $m_i \geq 1$ and  $z_i \in \bbC$.     The  general   element is a linear combination
of  finite sequences   $ : Z^1:   \cdots  : Z^r:  $  (possibly empty)   and has the form
\begin{equation}
F  =  \sum_{Z^1,  \dots  Z^r}   F(Z^1, \dots, Z^r) : Z^1:   \cdots  : Z^r:
\end{equation}

 Let     $\Up^w_{0}$  be the subset  with no  coinciding points between Wick products;  within Wick products  points may coincide.   For   $F \in  \Up_0^w$    we can define   an expectation 
$<F>$  by  
\begin{equation}  \label{prince}
< : Z^1:   \cdots  : Z^r:> =
 \sum_{P'}   \prod_{ \{(i,\al),(j, \beta)\} \in  P' }   C(m^i_{\al},z^i_{\al},m^j_{\beta},z^j_{\beta}   ) 
 \end{equation}
 This is analytic in  $z^i_{\al}  \in \bbC$  with  $z^i_{\al}  \neq z^j_{\beta}$ for   $i \neq j$.

Let   $\Up^w_{00}$   be elements   of   $\Up^w_0$  in which  points  within the Wick groups
do not coincide,    so there are no coinciding points at all. 
For  such  elements   the expectation agrees with
the expectation on    $\Up'_0$,  that is  
\begin{equation}  \label{uno}
< : Z^1:   \cdots  : Z^r:>      =    < : Z^1:_0   \cdots  : Z^r:_0>
 \end{equation}

 Now   consider
the  homomorphism   from    $ \al$  from  $ \Up^w_{00} $  to   $ \Up'_{0}$   defined  by  
$\al (:Z: )  =  :Z:_0$  that is  
\begin{equation}
 \al \Big(  : [ m_1,z_1, \dots,m_n, z_n]: \Big)    =       : [ m_1,z_1, \dots,m_n, z_n]:_0
\end{equation}
This is   onto    since   $:  [m_1,z_1]:  \cdots  :[m_n,z_n]: $  is sent to   $ [m_1,z_1,  \dots,   m_n,z_n] $.
It   is not  one-to-one  since for  example   if  $z_1 \neq  z_2$
both     $: [1, z_1,1,z_2]:  $    and  
 $ : [1, z_1]:  :[1,z_2]:  - C(z_1,z_2)  $     are  sent  to   $ :[1,z_1, 1,z_2]:_0$.
 By   (\ref{uno})   for   $F   \in    \Up^w_{00} $
 \begin{equation}
 < \al( F )   >   =  <  F   >
 \end{equation}

We   define    $J$  on     $  \Up^w $  by   the formula  (\ref{ooo})  and then   $\al \circ  J   =  J  \circ \al$.
For  $F  \in    \Up_0^w$   we have   $<JF>  =<F>$ directly from   (\ref{prince}).   We  also    define 
   $\Theta$  on     $  \Up^w_{\bbC - \{0\}} $  by   the formula  (\ref{ppp}).  Then    $\al \circ  \Theta   =  \Theta  \circ \al$
   and  $\Theta^2 = I$.
Furthermore  define a    linear mapping     $\nu:  \Up^w_{00,D-\{0\}}   \to  \cH$  by      $\nu =  \nu_0  \circ \al$
and then    for   $F,G    \in   \Up^w_{00,D-\{0\}}$  we have    
$   ( \nu(F),   \nu(G)   ) =<  \Theta( F  )   G  >  $.
We note in particular that      $\nu$ is defined so 
\begin{equation}  \label{85}
 \nu(  : [ m_1,z_1, \dots,m_n, z_n]: )    =  \nu_0(     : [ m_1,z_1, \dots,m_n, z_n]:_0)
\end{equation}

\bigskip

All this is just a reformulation of what we  already had.  But now we extend to allow coinciding points within
the Wick  product.   

\begin{lem}  The    mapping  $\nu:  \Up^w_{00,D-\{0\}}   \to  \cH$
extends  to a linear map   $\nu:  \Up^w_{0,D-\{0\}}   \to  \cH$
such that   for  $F,G   \in \Up^w_{0,D-\{0\}}$ 
\begin{equation}
   ( \nu(F),   \nu(G)   ) = <  \Theta( F )G  > 
\end{equation}
\end{lem}

\pr    Each  sequence  $Z=: [ m_1,z_1, \dots,m_n, z_n]:   \in  \Up^w_{0, D- \{0\}}$   can be  approximated by  a
a   sequence     $Z_k=: [ m_1,z_{1,k}, \dots,m_n, z_{n,k}]:   \in  \Up^w_{00, D-\{0\}}$ 
with  non-coinciding points  such that  $z_{j,k}  \to   z_j $   as   $k\to  \infty$.
Then   
\begin{equation}
\nu  (    : Z^1:   \cdots  : Z^r:  )   \equiv  \lim_{k \to \infty}  \nu(     : Z^1_k:   \cdots  : Z^r_k:    )
\end{equation}
exists   and is  independent of the approximating sequence.
This follows   from the continuity of the correlation functions.   This  defines   $\nu$  on  a basis
for    $ \Up^w_{0, D- \{0\}}$  and  hence   as a linear function on the whole algebra.    In the same way any     $F  \in   \Up^w_{0, D- \{0\}}$
can be  approximated   by   a sequence  $F_k   \in    \Up^w_{00, D- \{0\}}$  such that
$\nu  (  F )  =  \lim_{k \to   \infty}  
\nu (   F_k )  $.       Taking the limit   of   $ ( \nu(F_k),   \nu(G_k)   ) = <  \Theta(F_k)  G_k  >$
we  get   $  (\  \nu(F),   \nu(G)   ) = <  \Theta( F )G  > $.

 \subsection{}    
Next  we would like to  include  the  point  $z=0$  in  our analysis,  that is work with     $\Up^w_{0,D}$ 
 rather   than    $\Up^w_{0,D- \{0\}}$.    First   we  have

\begin{lem}  {   \  } \label{song}      Let  $z_i,  w_j  \in  D-\{0\}  $.  Then  
\begin{equation}  \label{oranges}
\begin{split}
&\Big( \nu (:[m_1,z_1, \dots  m_n, z_n]:),\nu ( : [\ell_1,  w_1, \dots,\ell_r, w_r]: )   \Big)\\
=& \pa^{m_1-1}_{\bar z_1}  \cdots   \pa^{m_n-1} _{\bar  z_n}   \pa^{\ell_1-1}_{  w_1}  \cdots   \pa^{\ell_r-1} _{ w_r} 
\left(  \sum_{  P' }  \prod_{  \{ \al, \beta \} \in  P'  }    ( 1  - \bar  z_{\al} w_\beta )^{-2}\right)\\
 \end{split}
\end{equation}
where the sum  is over pairings  $P'$  of  $(1, \dots,n)$  with  $(1, \dots, r)$  (empty if  $r \neq n$).
\end{lem}
\bigskip

\pr   If all the $m_i=1$   we  compute
\begin{equation} 
\begin{split}
\Big( \nu  ( : [ z_1, \dots, z_n]:),\nu  ( : [ w_1, \dots, w_r]:  )\Big)
= &    <( \Theta   : [ z_1, \dots, z_n]: ) : [ w_1, \dots, w_r]:  >  \\
=&      \prod_i  \frac{-1}{\bar z_i^2} \left<: \bar z_1^{-1}, \cdots, \bar z_n^{-1}:\  : w_1, \dots,  w_r:   \right>\\
=&      \prod_i   \frac{-1}{\bar z_i^2}  \sum_{P'} \prod_{ \{ \al, \beta\}  \in  P' }   C(  \bar z_{\al}^{-1},w_{\beta})\\
  =&  \sum_{  P' }  \prod_{  \{ \al, \beta\}  \in  P'  } \frac12   ( 1  - \bar  z_{\al} w_{\beta} )^{-2}\\
\end{split}
\end{equation}
The  result now follows  since by  (\ref{Wick}) 
\begin{equation}  
\begin{split}
&\Big( \nu (:[m_1,z_1, \dots  m_n, z_n]:),\nu ( : [\ell_1,  w_1, \dots,\ell_r, w_r]:  )   \Big)\\
=& \pa^{m_1-1}_{\bar z_1}  \cdots   \pa^{m_n-1} _{\bar  z_n}   \pa^{\ell_1-1}_{  w_1}  \cdots   \pa^{\ell_r-1} _{ w_r} 
 \Big(\nu  ( : [ z_1, \dots, z_n]:) ,\nu  ( : [ w_1, \dots, w_r]:  )  \Big)\\
 \end{split}
\end{equation}
\bigskip

\begin{lem}
The  mapping  $\nu:  \Up^w_{0,D-\{0\}}   \to  \cH$
extends  to a linear map   $\nu:  \Up^w_{0,D}   \to  \cH$
such that   for    $F  \in   \Up^w_{0,D}$   and     $G  \in    \Up^w_{0,D-\{ 0 \}}  $   
\begin{equation}  \label{fff}
   (\  \nu(G),   \nu(F)   ) = <  \Theta( G )F  > 
\end{equation}
\end{lem}

\pr   Each  sequence  $:Z:=: [ m_1,z_1, \dots,m_n, z_n]: $  with  $z_i \in D$  can be  approximated by  a
a   sequence     $:Z_k:=: [ m_1,z_{1,k}, \dots,m_n, z_{n,k}]: $  with  $z_{i,k}  \in   D- \{ 0\}$.  Then   
$\nu (: Z : )  \equiv  \lim_{k \to \infty}   \nu(  :Z_k:  )  $  exists  
 and is  independent of the approximating sequence.
This  follows from  the  representation  (\ref{oranges}).

More  generally  suppose    $  : Z^1:  :Z^2: \cdots  : Z^r:  \in  \Up^w_{0, D}$   has zeros in one factor,  say  $ :Z^1:$.
This    can be  approximated  by    $  : Z^1_k: :Z^2:  \cdots  : Z^r:  \in  \Up^w_{0, D- \{ 0\}}$  and  then 
\begin{equation}
\nu(   : Z^1:  :Z^2: \cdots  : Z^r:  )  \equiv  \lim_{k \to \infty}  \nu  (  : Z^1_k: :Z^2:  \cdots  : Z^r: )
\end{equation}
exists and   is  independent of the approximating sequence.  This  defines   $\nu$  on  a basis
for   $ \Up^w_{0,D}$   and  hence   as a linear function on the whole algebra. 
In the same   way   any    $F \in    \Up^w_{0, D}$  can be approximated  by  a  sequence  $F_k  \in  
 \Up^w_{0,D- \{0\}} $  such that  $\nu  (  F )  =  \lim_{k \to   \infty}  
\nu (   F_k )  $.    Then   $  (\nu(G),   \nu(F)   ) = <  \Theta(G)F  > $   follows by    taking
the limit of    $  (\nu(G),   \nu(F_k)   ) = <  \Theta(G)F_k  > $.

\begin{cor}  \label{snug2}
For  Wick  products  $ : Z^i:   =  :  [ m^i_1 z^i_1 \dots ,  m^i_{n_i} z^i_{n_i}]:$ 
the   function  $\nu (   : Z^1:  \cdots    : Z^r: )$  is strongly  analytic  in  $z^i_{\al}  \in  D,   z^i_{\al}  \neq  z^j_{\beta}$ for 
$i \neq  j$.    It is also  symmetric  under permutation of the $:Z^i:$  and under permutations within the $:Z^i:$.
\end{cor}
\bigskip

\pr  As in  corollary  \ref{snug}   $(\nu(F),\nu (   : Z^1:  \cdots    : Z^r: ))$  is analytic for  all  
$F \in   \Up^w_{0, D- \{ 0\}}$.    Such  $\nu(F)$  form a dense set,  hence  $\nu (   : Z^1:  \cdots    : Z^r: )$ 
is weakly analytic, and hence strongly analytic.  The symmetry follows similarly.

\newpage

\section{Creation and annihilation operators} 
\label{six}

Next we  define creation and  annihilation operators.        

\begin{thm}
For  $m \in \bbZ$  there exist   operators  $\al_m$  defined on  $  \nu( \Up^w_{0,D - \{ 0\}})$   such that  
\begin{equation}  \label{ping}
\al_m   \nu ( F  )  =   \sqrt{2} \oint_{|z|=r }  \frac{dz}{2\pi}\  z^m \  \nu(  [1,z] F )   
\end{equation}    with  $0<r<1$  selected to     
 enclose  all  the points  in  $F$.  (By the analyticity the expression  is
 independent of the   choice  of  $r$.)
Furthermore  with   $\Om = \nu(\emptyset)$
 \begin{enumerate}
 \item   $(\al_m)^*  =  \al_{-m}$
 \item   $\al_m  \Om =0$   for   $m  \geq  0$.
  \item   $ [ \al_m,  \al_n]  =  m  \de_{m+n}  $
 \end{enumerate}
\end{thm}
\bigskip

\re  These are standard  arguments which we adapt to our setup.
\bigskip
 
\pr   Abbreviating  $[1,z]$  as   $[z]$  we    first   define  $ \al_m$  on    $ F   \in   \Up^w_{0,D-\{0\}}  $ by    
\begin{equation}
 \al_m  F    =   \sqrt{2}  \oint _{|z|=r}  \frac{dz}{2\pi}\  z^m \  \nu(  [z]F )   
\end{equation}  
Then  making the change of variables   $z = \bar w^{-1},  dz   = - \bar w^{-2}  d \bar w$ we have 
\begin{equation}  \label{pong} 
\begin{split}
(\nu(G),   \al_m F )   = &   \sqrt{2}  \oint _{|z|  =r}  \frac{dz}{2\pi}\  z^m \   \Big(\nu (G),  \nu(  [z]F  ) \Big)  \\
 = &   \sqrt{2}  \oint _{|z|  =1}  \frac{dz}{2\pi}\  z^m \   < (\Theta  G)[z] F >  \\
  = &   \sqrt{2}  \oint _{|w|  =1}  \frac{d   \bar w  }{2\pi}\  (-\bar w^{-m-2} )  \   < (\Theta G) [\bar w^{-1}]  F >  \\
    = &   \sqrt{2}  \oint _{|w|  =1}  \frac{d   \bar w  }{2\pi}\  \bar w^{-m}   \   < (\Theta [w] G) F >  \\
   = &   \sqrt{2}  \oint _{|w|  =r}  \frac{d \bar w}{2\pi}\  \bar w^{-m} \  \Big( \nu( [ w] G),  \nu( F )\Big)  \\
    = &   (\al_{-m} \nu( G),  \nu(F))  \\
\end{split}
\end{equation}
Now    if   $\nu(F)   = 0$   then   $(\nu(G),   \al_m F )=0$  for  all   $\nu(G) $  and hence   $\al_m F  =  0$.
Thus  we  can   define  $\al_m$   on   $  \nu( \Up^w_{0,D - \{ 0\}})$   by
$ \al_m   \nu (F)   =  \al_m  F$.  Then  (\ref{ping})   holds   and   (\ref{pong})  can be written    
\begin{equation}
 (\nu(G),   \al_m  \nu (F ) = (\al_{-m}   \nu( G),  \nu(F ))
\end{equation}
which  says    $(\al_m)^*  =  \al_{-m}$.

The    result    $\al_m  \Om =0$   for   $m\geq 0$   follows by analyticity.

For the third  item     we  compute
 \begin{equation}
 \begin{split}
 (\nu(G),   [ \al_m,  \al_n]  \nu(F) )   
 = &2 \int_{\Ga^+-\Ga_-}  \frac{dz}{2\pi}   \oint_{|\zeta| =r}  \frac{d\zeta  }{2\pi} z^m  \zeta^n 
  <(\Theta  G)[z, \zeta ]  F > \\
\end{split}
\end{equation}
where   $\Ga^{\pm}$  are the contours  $|z|=r \pm \ep$ and  $\ep$ is sufficiently small. 
Now we claim that  
\begin{equation}
  <(\Theta  G)[z, \zeta ]  F >=   C(z, \zeta)  <(\Theta  G)  F >
  +   \cdots
\end{equation}
where  the  term  $\dots$  is analytic  in  $z$  between  $\Ga^{\pm}$.
To see this recall that  
the  correlation functions  are  written    as  sums over 
pairings.   If    the points  $z, \zeta$  are not paired with each other  then they are paired with
something  outside of   the corridor between  $\Ga^{\pm}$ and hence   the expression
is analytic between   $\Ga^{\pm}$.  The terms where they are paired with
each other have the  claimed  form.
In the integral the analytic term  contribute nothing.  For the other term we compute
\begin{equation}
   \oint_{|\zeta| = r}  \frac{d\zeta  }{2\pi}   \int_{\Ga^+-  \Ga^-}  \frac{dz}{2\pi} 
     \frac{- z^m  \zeta^n}{(z-\zeta)^2}  
=           \oint_{|\zeta| = r}  \frac{d\zeta  }{2\pi} (- im) \zeta^{n+m-1} =  m  \de_{m+n}  
\end{equation}
Thus we  get 
\begin{equation}
\Big(\nu(G),   [ \al_m,  \al_n] \nu(F) \Big)    =   m  \de_{m+n}  \Big(\nu(G),  \nu(F) \Big) 
 \end{equation}
as required.  This  completes the proof.
\bigskip

Now  consider   vectors    $\nu  (   : [z_1, \dots, z_n]:  )  \equiv  \nu  (   : [1,z_1, \dots,1, z_n]:  )$.  By  
 Corollary  \ref{snug2}  this is analytic    in  $z_i \in D$   and  hence has a convergent power 
 series in this region.  The next  result  identifies the coefficients   in the  series  

\begin{lem}   For     $z_i \in D$  
  \begin{equation}  \label{pretty}
  \nu  (   : [z_1, \dots, z_n]:  )
= \left(\frac   { 1}{\sqrt 2 i} \right)^{n/2}   \sum_{m_1,  \dots ,  m_n \geq  0 } z_1 ^{m_1}  \dots   z_n^{m_n} \  \Big( \al_{-m_1-1}  \cdots  \al_{-m_n-1} \Om \Big)
\end{equation}
or  
 \begin{equation}  \label{stinger}
  \nu  (   : [z_1, \dots, z_n]:  )
= \left(\frac   { 1}{\sqrt 2 i} \right)^{n/2}   \sum_{m_1,  \dots ,  m_n \geq  1 } z_1 ^{m_1-1}  \dots   z_n^{m_n-1} \  \Big( \al_{-m_1}  \cdots  \al_{-m_n} \Om \Big)
\end{equation}
\end{lem}

\pr       The coefficient  of         $ z_1 ^{m_1}  \dots   z_n^{m_n} $   is  
\begin{equation}
\int_{|z_n|=r_n}  \frac{ dz_n}{2\pi i} \frac{1}{  z_n^{m_n+1} } \cdots   \int_{|z_1|=r_1}   \frac{ dz_1}{2\pi i}
  \frac{1}{  z_1^{m_1+1} }  \      \nu  (   : [z_1, \dots, z_n]:  )
\end{equation}
for any   $0 < r_i <1$  and we take   $r_1 <  r_2<  \cdots  < r_n$.
But for    distinct  $z_i$ 
\begin{equation}
\begin{split}
&\nu  (   : [z_1, \dots, z_n]:  ) 
=    \nu  ( [z_1]  : [z_2, \dots, z_n]:  ) 
-     \sum_{j=1}^n   C(z_1,z_j)
\nu  (   : [z_1, \dots, \widehat{z_j},  \dots,  z_n]:  )  \\
\end{split}
\end{equation}
   This is true  with  $\nu_0  (   : [z_1, \dots, z_n]:_0  ) $  by  (\ref{ugh})   and hence also holds  as stated by  (\ref{85}).
The  second  term  is  analytic inside 
$|z_1| = r_1$  and so does not contribute  to  the integral.  For the first 
term  we have  
\begin{equation}
\int_{|z_1|=r_1}  \frac{ dz_1}{2\pi i} \  \frac{1}{  z_1^{m_1+1} } \  \nu  ( [z_1]  : [z_2, \dots,z_n]:  )   =  \frac{1}{\sqrt{2}i}  \al_{-m_1-1}  
 \nu  (  : [z_2, \dots,z_n]:  )   
\end{equation}
Repeating this argument  
\footnote{  We  are taking  the operators  $\al_m$  outside the integral without having established continuity.
It is allowed since it suffices to establish the identity weakly}
we find the  coefficient  of         $ z_1 ^{m_1}  \dots   z_n^{m_n} $   is   
\begin{equation}
\left(\frac   { 1}{\sqrt{2}i} \right)^{n}  \al_{-m_1-1}  \cdots   \al_{-m_n-1}  \Om 
\end{equation}

\begin{thm}  The vectors 
$ \al_{-m_1}  \cdots   \al_{-m_n}  \Om $   with  $m_i \geq  1$  span a  subspace    dense  in   $\cH$.   Hence   $\cH$ is separable
and    naturally
isomorphic to  Fock space.
\end{thm}
\bigskip

\pr 
Vectors  of the form    $ \nu_0    (    [z_1, \dots, z_n]  ) $ with  $z_i \neq z_j$ span a dense subspace  by  construction.  
But   $ [z_1, \dots, z_n]$  is a   combination of the 
 $:[z_1, \dots, z_n]:_0  $  (use the identity   (\ref{ugh}))  .  Hence  the    $ \nu_0    (   : [z_1, \dots, z_n]:_0  ) $ span a dense subspace 
and these are the same as  the   $ \nu   (   : [z_1, \dots, z_n]:  ) $.     Thus  it suffices to approximate
vectors    $ \nu   (   : [z_1, \dots, z_n]:  ) $   with vectors     $ \al_{-m_1}  \cdots   \al_{-m_n}  \Om $.
This follows by truncating the power series   (\ref{stinger}).
\bigskip

\re   We  note that  $\al_0 =0$.    This follows  on vectors  $ \al_{-m_1}  \cdots   \al_{-m_n}  \Om$
by  $[\al_0, \al_{-m} ]  =0$  and  $\al_0\Om =0$.
\bigskip
 
 \begin{lem}
\begin{equation}
  \al_{-m_1}    \cdots    \al_{-m_n }  \Om
= \prod_j \left( \frac{\sqrt{2} i }{(m_j-1)!}  \right)                  \ \   \nu  \Big(  :  [m_1,0, \dots,  m_n,0]   : \Big)
\end{equation}
or if  $n_m$  is the number of times  $m$ occurs in  $m_1,  \dots,  m_n$
\begin{equation}  \label{under}
  \al_{-m_1}    \cdots    \al_{-m_n }  \Om
= \prod_m \left( \frac{\sqrt{2} i }{(m-1)!}  \right)^{n_m}                  \ \   \nu  \Big(  :  \prod_m [m,0]^{n_m}   : \Big)
\end{equation}
\end{lem} 
\bigskip

\pr   We  compute  by (\ref{Wick})  and    (\ref{stinger})
\begin{equation}
\begin{split}
 \nu  \Big(  :  [m_1,0, \dots,  m_n,0]   : \Big)
=  &  \pa^{m_1-1}_{z_1}  \cdots   \pa^{m_n-1} _{z_n} 
 \nu  \Big(  :  [z_1, \dots, z_n]   : \Big)  \Big|_{z_i =0}\\
= &\prod_j \left( \frac{(m_j-1)!} {\sqrt{2} i } \right)      \    \al_{-m_1}    \cdots    \al_{-m_n }  \Om\\
\end{split}
\end{equation}
    \bigskip

\textbf{Summary}:   We have  constructed a Hilbert space $\cH$   from  the correlation functions for  the model.     In this  space  we  have constructed  a  vacuum $\Om$   and   creation and annihilation 
 operators   $\al_m$     such that    vectors    $ \al_{-m_1}    \cdots    \al_{-m_n }  \Om$  span  a dense set,    so 
$\cH$   is identified as a Fock space.    Furthermore    vectors    $ \al_{-m_1}    \cdots    \al_{-m_n }  \Om$
are identified  with    states    $ \nu (: [m_1,0, \dots,  m_n,0]: ) $  representing Wick products of  fields
at the origin.     These are the  basic facts  which we need to attack the main problem.

\section{Transition amplitudes}  \label{seven}

Now we define transition amplitudes.   Let     $D_1,  \dots,  D_r$  be  a number of  disjoint    discs 
in  $\bbC$.  
They    are   of the form
\begin{equation}
D_i  =  \{  z \in \bbC:      |z  -a_i|  < R_i  \}
\end{equation}    
A natural  problem is      to  consider a correlation function of the form 
$<G_1  \cdots  G_r>$ with  $G_i \in \Up_{D_i- \{a_i\}}$   and show that  it
defines a   bounded  multi-linear functional on  $\cH_{D_1}  \times  \cdots  \times   \cH_{D_n}$  where   $\cH_{D}$
is defined  in  (\ref{fifty}).
That is    we  would like to show that  it depends only   on  the equivalence classes  $\tilde \nu (G_i) \in  \cH_{D_i}$
and   that  
there is a constant  $C$  such that  
\begin{equation}
 |<G_1  \cdots  G_r>|   \leq   C   \| \tilde \nu(G_1)\|_{D_1}  \cdots      \| \tilde \nu(G_r)\|_{D_r}  
\end {equation}
Then   it  extends by continuity     to   $\cH_{D_1}  \times  \cdots  \times   \cH_{D_n}$.

It is     desirable    to  refer everything to the standard  Hilbert space  $\cH$ 
on the disc  $D= \{  z \in \bbC:   |z|  <1\}$.   
Accordingly  we    assume the  $D_i$   are  parametrized   by   mappings $j_i$  from  $D$ to  $D_i$
of the form
$ j_i(z)  =    a_i +    q_i  z   $  with      $|q_i|  =  R_i$.
As  we have seen in section  \ref{serious}   these  induce 
isomorphisms  $J_i$   from  $\Up_{0,D-\{0\}}$ to  $\Up_{0, D_i- \{a_i\}}$
and  unitary maps  $\cJ_i$ from 
$\cH$  to  $\cH_{D_i}$  such that  $\cJ_i  \nu_0(F)   = \tilde   \nu  (J_i F) $.
The  problem is reformulated    as  follows.  Given      $F_1, \dots,  F_r$  in  $\Up_{0,D-\{0\}}$
show that   $ <    (J_1F_1)  \cdots     (J_rF_r)>$   depends only  on  $\nu_0(F_i) \in \cH$  and  
that  
\begin{equation}
 |<(J_1F_1)  \cdots     (J_rF_r)>|   \leq   C   \|  \nu_0(F_1)\|  \cdots      \| \nu_0(F_r)\|  
\end {equation}
If  we  put   $F_i   =  J_i^{-1} G_i$  we get the previous version.

We   further generalize  by  allowing  Wick products.   If    $F_1, \dots,  F_r$  in  $\Up^w_{0,D}$  then  
   $J_iF_i  \in  \Up^w_{0,D_i}$  and we       consider    correlation functions    
$ <    (J_1F_1)  \cdots     (J_rF_r)>$.   We    seek to show  that these depend only on   $\nu(F_i)  \in   \cH$
and that  
\begin{equation}
 |<(J_1F_1)  \cdots     (J_rF_r)>|   \leq   C   \|  \nu(F_1)\|  \cdots      \| \nu(F_r)\|  
\end {equation}
Given  $F_i'  \in   \Up_{0,D-\{0\}}$   one can choose  $F_i  \in  \Up^w_{00,D-\{0\}}$  such that 
$\al(F_i)   = F'_i$    and hence recover the previous version.

Let  $\cD$ be the dense domain   $\cD  \equiv  \nu( \Up^w_{0,D} )$.
As  a first step    we  have   as in      \cite{Dim07}:

\begin{lem}
For    $F_i  \in  \Up^w_{0,D} $   the  correlation function  $ <    J_1F_1, \dots,   J_rF_r>$
only  depends   on      $\cF_i=\nu(F_i) \in  \cD$.   It thus     defines 
a multilinear functional $\cA_r$  on $  \cD  \times  \cdots   \times    \cD$  such that
\begin{equation}
\cA_r  (  \cF_1,  \cdots ,   \cF_r)   =    <   ( J_1F_1) \dots ( J_rF_r)>
\end{equation}
Furthermore if   $\phi_{\al} =  \nu ( \psi_{\al}) \in \cD$  is an orthonormal  basis  for  $\cH$  then  
\begin{equation}  \label{cream}
\cA _r (  \cF_1,  \cdots ,   \cF_r)   =  \sum_{\al_1}  \cdots   \sum_{\al_r}    
\cA _r (\phi_{  \al_1},  \cdots ,\phi_{   \al_r})  (\phi_{\al_r},  \cF_r) \cdots    (\phi_{\al_1},  \cF_1)
\end{equation}
\end{lem}
\bigskip

\re   We  do not yet assert that  the functional  is  bounded.  We  also   do  not yet assert  that the multiple sum is absolutely  convergent,  only that the iterated
sum converges.  
\bigskip

\pr  
 Since  $<J F> = <F>$ we have 
\begin{equation}
 <   ( J_1F_1)(J_2F_2) \dots  ( J_rF_r)>
=   <   F_1(J_1^{-1}J_2F_2) \dots (  J_1^{-1} J_rF_r)>
\end{equation}
Since the  $D_i$  are  disjoint we  have  for  $i \neq 1$ that    $j_iD  =D_i \subset D_1'$, hence      $j_1^{-1}j_i D  \subset  D'$,
and hence    $J_1^{-1}  J_i  F_i  \in   \Up^w_{0, D'}$.   Since  $\Theta^2 =I$   it   follows  that   $( J_1^{-1} J_2F_2) \dots (  J_1^{-1} J_nF_n)$
has the form  $\Theta F$   for some   $F  \in   \Up^w_{0, D-\{0\}}$  and so    by  (\ref{fff})
\begin{equation}  \label{stay}
 <   ( J_1F_1)(J_2F_2) \dots  ( J_rF_r)>
=   <   (\Theta   F) F_1 >  =   (\nu(F),  \nu(F_1))
\end{equation}
Thus   the  expression only depends  on  $\cF_1= \nu(F_1)$.  A similar argument establishes the
result in the other variables  and hence the   first   result.

To  get the expansion   we  compute  from  (\ref{stay})  
\begin{equation} 
\begin{split}
\cA_r(\cF_1, \dots,  \cF_r)  
= &  \sum_{\al_1}    (\nu(F), \phi_{\al_1})( \phi_{\al_1},      \nu(F_1))\\
= &  \sum_{\al_1}    <  \Theta(F)\psi_{\al_1}>\   ( \phi_{\al_1},  \nu(F_1))\\
= &  \sum_{\al_1}   <   ( J_1  \psi_{\al_1})(J_2F_2) \dots  ( J_rF_r)>( \phi_{\al_1},   \cF_1)\\
= &  \sum_{\al_1}  \cA_r( \phi_{\al_1},  \cF_2, \dots,  \cF_r)    ( \phi_{\al_1},   \cF_1)\\
\end{split}
\end{equation}
Repeating this  in the other variables gives the result.
\bigskip

Now the  question is whether   $\cA_r$  extends  from   a functional on 
the dense  domain   $\cD  \times   \cdots  \times  \cD$   to the full Hilbert space
$\cH  \times   \dots   \times  \cH$.  
 We  prove the stronger result   that  
it is a Hilbert-Schmidt functional, i.e.  that there is a orthonormal basis   
$\{\phi_{\al}\}    $  for  $\cH$ with   $\phi_{\al}  \in \cD$   such that
\begin{equation}
\|  \cA_r  \|^2_{HS}   =  \sum_{\al_1,   \dots,  \al_r  }    | \cA _r (  \phi_{\al_1},   \cdots        \phi_{\al_r})|^2 
\  < \infty
\end{equation}
If   it converges for one orthonormal basis then it converges for  all such bases.
Then  by the Schwarz inequality the sum   (\ref{cream})  is absolutely convergent and 
we have the bound   for   $\cF_i \in \cD$
 \begin{equation}  \label{toot}
|\cA_r  ( \cF_1, \dots   , \cF_r) |     \leq   
\|\cA_r\|_{HS}    \|  \|\cF\|_1   \cdots  \| \cF\|_n
\end{equation}
Hence    $\cA_r$  extends to a bounded  multilinear functional  on  $\cH \times   \dots   \times  \cH$  which  is still    Hilbert-Schmidt.

To  state the result   precisely    define
\begin{equation}
R  =  \sup_i  R_i   \hspace{1cm}      d  =  \inf_{i,j}  |a_i-a_j|
\end{equation}
We  assume that  $d/R$ is not too small.   Note  that 
the  Euclidean  distance  between  $D_i$ and $D_j$  is  
\begin{equation}
d(D_i,D_j)   =  |a_i-a_j| - R_i-R_j  \geq  d -2R
\end{equation}     Hence    $d/R >2$  guarantees
that   the discs are separated.

\begin{thm}  
Let   $d/R >  4 \sqrt {r} \geq 4$.  Then   $\cA_r$ on $\cD  \times   \cdots  \times  \cD$   is a Hilbert-Schmidt  functional  
and  so   extends to  a bounded multilinear  functional  on  $\cH  \times   \dots   \times  \cH$
satisfying  (\ref{toot}).  
\end{thm}
\bigskip

\pr   An orthonormal  basis $\phi(  \{ n_m  \} )  $   for Fock space is  indexed  by sequences  $\{ n_m\}  = \{n_1,n_2, \dots \}$  which are 
eventually  zero.  It   has the form    
\begin{equation}
\phi(  \{ n_m  \} )  =   \prod_m (n_m!  m^{n_m})^{-1/2}  \prod_m   \al_{-m}^{n_m} \Om
    \end{equation} 
 Therefore by  (\ref{under}) an orthonormal basis 
for  $\cH$   is   
\begin{equation}  \label{basis}
\phi(  \{ n_m  \} )  =   \prod_m ( n_m! )^{-1/2} \left(\frac{ i\sqrt{2m}}{m!} \right)^{n_m}    \nu  (  :  [m_1,0, \dots,  m_n,0]   : )
\end{equation} 
Here    $\nu(  :  [m_1,0, \dots,  m_n,0]   : )$   has  $n^m$  $m'$s  in any order  and can   also be written
 $ \nu(  :  \prod_m[m,0]^{n_m}    : )$.   We  also define  in  $\Up^w_{0,D}$
  \begin{equation}  \label{basis2}
\psi(  \{ n_m  \}, z )  =   \prod_m ( n_m! )^{-1/2} \left(\frac{ i\sqrt{2m}}{m!} \right)^{n_m}    :  [m_1,z, \dots,  m_n,z]   : 
\end{equation} 
and then    $ \phi(  \{ n_m  \} )  =  \nu    (\psi(  \{ n_m  \}, 0 )  ) $

We  want to study   
 \begin{equation}
\cA_r  \Big( \phi( \{ n^1_m\}),       \dots,         \phi( \{ n^r_m\})\Big)  
=  \Big<  J_1  \psi\big( \{ n^1_m\},0\big)       \cdots      J_r  \psi \big( \{ n^r_m\},0\big)\Big >
\end{equation}
Since  
\begin{equation}
J_j  \psi(  \{ n^j_m  \}, 0 )
= \prod_m   q_j^{ n^j_m m}   \ \   \psi(  \{ n^j_m  \}, a_i )
\end{equation}   we
 have  
 \begin{equation}   \label{southern}
\cA_r  \Big( \phi( \{ n^1_m\}),       \dots,         \phi( \{ n^r_m\})\Big)  
=  \Big<   \psi\big( \{ n^1_m\},a_1\big)       \cdots       \psi \big( \{ n^r_m\},a_r\big)\Big> 
\prod_{j=1}^r   \prod_{m=1}^{\infty} q_j^{    n^j_m m} 
\end{equation}
By  (\ref{prince}):  
\begin{equation}   \label{northern}
\begin{split}
& \Big<   \psi\big( \{ n^1_m\},a_1\big)       \cdots      \psi \big( \{ n^r_m\},a_r\big)\Big> \\
=& \prod_{j,m}   ( n^j_m! )^{-1/2} \left(\frac{ i\sqrt{2m}}{m!} \right)^{n^j_m}   \sum_{P'}   \prod_{ \{(i,\al),(j, \beta)\}  \in  P'  } 
  C(m^i_{\al},a_i,m^j_{\beta},a_j   )\\
\end{split}
\end{equation}

Now we estimate these quantities.  Using  $(a+b)!  \leq  a!b!2^{a+b}$  and       (\ref{lady})
\begin{equation}
\big|C(m^i_{\al},a_i,m^j_{\beta},a_j   )\Big|   \leq \frac12  (  m^i_{\al}  +m^j_{\beta}-1)!   d^{-m^i_{\al}-m^j_{\beta}}
\leq \frac{1}{2^2}  m^i_{\al}!m^j_{\beta}!     \left( \frac{2}{d}  \right)^{m^i_{\al}+m^j_{\beta}}
\end{equation}
The    $m!$'s   contribute  a   factor    $\prod_{j,\beta} m^j_{\beta}!  =  \prod_{j,m}   (m!)^{n^j_m}$   which is  canceled by a  similar factor 
in  (\ref{northern}).
  The factors  
$(2/d)^{m^i_{\al}+m^j_{\beta}}$  combine to give an overall  $ \prod_{j,\beta}  (2/d)^{ m^j_{\beta}}= \prod_{j,m}  (2/d)^{ n^j_m m} $.
The  factors   $2^{-2}$   contribute a factor  $2^{-n}$   where  
\begin{equation}
n =\sum_{j=1}^r  \sum_{m=1}^{\infty}   n^j_m
\end{equation}
is the total number of fields.    
Also we   estimate the number of  pairings    or  graphs    by ignoring   the restriction that a line cannot
join  the same vertex.   Thus it is fewer than the number of   graphs on  $n$  legs which   
 is   
\begin{equation}
 \frac{ n!}{  (n/2)! 2^n }   \leq   \sqrt{ n!} 2^{-n/2}
 \end{equation}
 and the  $2^{-n/2}$  is canceled by   a similar factor  in  (\ref{northern}).
 Altogether  then   we have  
\begin{equation}
\left| \Big<   \psi\big( \{ n^1_m\},a_1\big)       \cdots      \psi \big( \{ n^r_m\},a_r\big)\Big>  \right|
\leq   \sqrt{ n! } 2^{-n}
  \prod_{j,m} ( n^j_m! )^{-1/2}
m^{n^j_m/2} \left(  \frac{2}{d}  \right)^{ n^j_m m}  
\end{equation}
We  take  $m<2^m$  in this and  combine it with the estimate
\begin{equation}
 \prod_{j,m} |q_j|^{  n^j_m m}  \leq    \prod_{j,m} R^{    n^j_m   m}  \end{equation}
to obtain   
\begin{equation}
\left| \cA_r  \Big( \phi( \{ n^1_m\}),       \dots,         \phi( \{ n^r_m\})\Big)  \right|   \leq   
  \sqrt{ n!}\ 2^{-n}
   \prod_{j,m} ( n^j_m! )^{-1/2}    \left( \frac{2^{3/2}R}{d}  \right) ^{  n^j_m m}  
\end{equation}

We   want to  sum  $|\cA_r|^2$   over  all  integers    $n^j_m \geq   0$  with 
  $1 \leq j \leq  r$  and    $ 1 \leq  m $. 
  We  first  consider a smaller sum with the additional  restriction   $  m \leq  M $.    Then   by the multinomial expansion:
      
\begin{equation}    \label{bigsum}
\begin{split}
\sum_{\{\{n^j_m\}:  m\leq  M\}}    |\cA_r |^2 
 \leq &  \sum_{\{\{n^j_m\}:  m\leq  M\}}   n!\   4^{-n}
  \prod_{j,m} ( n^j_m! )^{-1}
  \left(   \frac{8R^2}{d^2}  \right)  ^{  n^j_m m}  \\
=    &      \sum_{n=0}^{\infty}    \  4^{-n} 
 \sum_{ \{\{n^j_m\} :   m\leq  M,    \sum n^j_m =n\} }  \frac{n!}
{  \prod_{j,m}  n^j_m! }
 \left[ \left(   \frac{8R^2}{d^2}  \right)  ^ m  \right]^{n^j_m}  \\
=     &      \sum_{n=0}^{\infty}     4^{-n} 
 \left(  \sum_{j=1}^r   \sum_{m=1}^M
 \left(   \frac{8R^2}{d^2}  \right)  ^ m \right)^n \\
=     &      \sum_{n=0}^{\infty}    
 \left(   \frac{r}{4} \  \sum_{m=1}^M 
 \left(  \frac{8R^2}{d^2}  \right)  ^{   m} \right)^n \\
\end{split}
\end{equation}
However  the sum over   $m$  is dominated by the infinite sum and  
\begin{equation}
  \frac{r}{4} \sum_{m=1}^{\infty} 
 \left(  \frac{8R^2}{d^2}  \right)  ^{   m}   =    \frac{r}{4} \     \frac{8R^2/d^2}{  1-   8R^2/d^2}   \  <    \frac{1}{4} 
\end{equation}
since under   our  assumption   $8R^2/d^2   \leq  8rR^2/d^2  <  1/2  $.   Then  the final    sum   over $n$  in  (\ref{bigsum})  converges    and is bounded in    $M$.   
 Hence  the  sum  $\sum_{\{n^j_m\}}  |\cA_r |^2 $ converges without the restriction    $m \leq  M$    which is our result.
  \bigskip

\res
To  extend these results to bosonic string theory  and define   (tree level, genus zero) string theory scattering amplitudes  one
would  have    to make a number of  modifications  which we now list.

\begin{enumerate}  
\item   Replace  $X: \bbC_{\infty}  \to  \bbR$   by  $X: \bbC_{\infty}  \to  \bbR^d$
where  $d$ is the dimensional of spacetime,  preferably $d=26$.  The basic
covariance is then
\begin{equation}
<   \pa X^{\mu} (z)  \pa X^{\nu}(z')>   =   -   \frac12 \de^{\mu \nu}  (2\pi)^{-1}(z-z')^{-2}  
\end{equation}

\item  Replace the monomials  $ :\pa^{m_1}X^{\mu_1}(z)  \cdots   \pa^{m_n}X^{\mu_n}(z):$
by expressions (vertex functions)
\begin{equation}
 :\pa^{m_1}X^{\mu_1}(z)  \cdots   \pa^{m_n}X^{\mu_n}(z) e^{ik\cdot  X(\bar z, z)}:
\end{equation}
The factor  $ e^{ik\cdot  X(\bar z, z)}$  with  $k \in  \bbR^d$  is present to accommodate the 
center of mass motion.

\item   Analytically continue   from the Euclidean
metric $ \de_{\mu \nu}  $  to  the  Minkowski metric
$ \eta_{\mu \nu}$.  Show that the correlation functions  define  multilinear functionals  on a Hilbert
space with a  particle interpretation.  This will require  that   the momenta  $k$  be 
on the mass shell  and  has   to deal with the fact   that the natural inner products  with 
the Minkowski metric 
are indefinite.    

\item   Integrate over the positions  of   the vertex functions  over  $\bbC_{\infty}$.
This will involve isolating the singularities that occur.
 \end{enumerate}

This program is reasonably  well understood for  a  few low lying states  (see  Polchinski  \cite{Pol98}),
but a  systematic treatment  is  lacking

\appendix


\begin{thebibliography}{99}



\bibitem{Dim04}  J. Dimock,  Markov quantum fields on a manifold, Rev. Math. Phys \textbf{16},
243-255, 2004


\bibitem{Dim07}  J. Dimock,  Transition amplitudes and sewing properties for bosons on the Riemann
spere,  J.  Math. Phys.   \textbf{48},  052308, 1-31,  2007.




\bibitem{FFK89}  G. Felder, J. Frohlich, J. Keller, On the structure
of unitary conformal field theory, Commun. Math. Phys. \textbf{124}, 417-463, 1989




\bibitem{Gaw99}  K. Gawedski, Lectures on conformal field theory,  in {\em Quantum fields and strings: a course for 
mathematicians},   P. Deligne et. al. eds., American Mathematical Society, Providence, 1999. 








\bibitem{Pic07}  D. Pickrell,   $P(\phi)_2$ quantum field theories and Segal's axioms.  (math-ph/0702077)

\bibitem{Pol98}  J. Polchinski,   String Theory,   Cambridge University Press,  Cambridge,  1998.



\bibitem{Sch97}   M.  Schottenloher,  {\em A mathematical introduction to conformal 
field theory},   Springer,  1997







\end{thebibliography}
\end{document}